\documentclass[nonacm,sigconf]{acmart}

\usepackage{booktabs}
\usepackage{alltt} 
\usepackage{geometry} 
\usepackage{tikz} 
\usepackage{balance}
\usepackage{xcolor}
\usepackage{subcaption}
\usepackage{float}
\usepackage{ulem}

\newcommand*\circled[1]{\textcircled{\small {#1}}}

\begin{document}
\title{ACE: Towards Application-Centric Edge-Cloud Collaborative Intelligence}

\begin{abstract}
Intelligent applications based on machine learning are impacting many parts of our lives.
They are required to operate under rigorous practical constraints in terms of service latency, network bandwidth overheads, and also privacy.
Yet current implementations running in the Cloud are unable to satisfy all these constraints.
The Edge-Cloud Collaborative Intelligence (ECCI) paradigm has become a popular approach to address such issues, and rapidly increasing applications are developed and deployed.
However, these prototypical implementations are developer-dependent and scenario-specific without generality, which cannot be efficiently applied in large-scale or to general ECC scenarios in practice, due to the lack of supports for infrastructure management, edge-cloud collaborative service, complex intelligence workload, and efficient performance optimization.
In this article, we systematically design and construct the first unified platform, ACE, that handles ever-increasing edge and cloud resources, user-transparent services, and proliferating intelligence workloads with increasing scale and complexity, to facilitate cost-efficient and high-performing ECCI application development and deployment.
For verification, we explicitly present the construction process of an ACE-based intelligent video query application, and demonstrate how to achieve customizable performance optimization efficiently.
Based on our initial experience, we discuss both the limitations and vision of ACE to shed light on promising issues to elaborate in the approaching ECCI ecosystem.
\end{abstract}

\author{Luhui Wang}
\affiliation{
\institution{Xi'an Jiaotong University}
}

\author{Cong Zhao}
\affiliation{
\institution{Imperial College London}
}

\author{Shusen Yang}
\affiliation{
\institution{Xi'an Jiaotong University}
}

\authornote{S. Yang (Corresponding Author) is with Xi'an Jiaotong University and Pazhou Laboratory. Email: shusenyang@mail.xjtu.edu.cn}

\author{Xinyu Yang}
\affiliation{
\institution{Xi'an Jiaotong University}
}

\author{Julie McCann}
\affiliation{
\institution{Imperial College London}
}

\maketitle

\section{Introduction} \label{sec:introduction}
In recent years, machine learning, especially deep learning, has been applied to various domains (\textit{e.g.}, computer vision, speech recognition, and video analytics).
Emerging \textit{Intelligent Applications} (IAs) such as image classification based on deep Convolutional Neural Networks (CNNs)~\cite{Krizhevsky2017}, traffic flow prediction based on deep Recurrent Neural Networks (RNNs)~\cite{WangZhumei2020}, and game development based on deep Generative Adversarial Networks (GANs)~\cite{Kim2020}, are demonstrating superior performance in terms of accuracy and latency.
Such performance, however, requires tremendous computation and network resources to deal with the increasing size of Machine Learning (ML)/Deep Learning (DL) models and the proliferation of vast amounts of training data~\cite{MayerJ2020}.

Cloud computing is indisputably attractive to IA developers as the predominating high-performance computing paradigm~\cite{10.1145/3364684}.
Typically, cloud providers offer services like Infrastructure-as-a-Service (IaaS), Platform-as-a-Service (PaaS), and Software-as-a-Service (SaaS) to facilitate application implementation, where resources like high-performance computation, massive elastic storage, and reliable network services are allocated according to user requirements.
Intuitively, mainstream IAs are deployed on the Cloud to leverage centralized resources for computationally-intensive Artificial Intelligence (AI) tasks like data processing, ML/DL model training, and inference.
For instance, the distributed training of AlphaGo~\cite{Silver2016} is a typical \textit{`Cloud Intelligence'} (CI) representative.

However, novel challenges to CI emerge when modern IAs rapidly proliferate and are required to be in production in practice, where \textit{high end-to-end service latency}, \textit{high network bandwidth overhead}, and \textit{severe privacy leakage threat} are among the most critical ones~\cite{zhou2019edge}.
Instead of concentrating on the Cloud, increasing efforts attempt to exploit heterogeneous resources distributed at the network Edge to address such issues.
For example, some IAs offload DL tasks to edge servers (\textit{e.g.}, Nvidia Jetson TX2 Board)~\cite{Daniel2019} for privacy preservation and timely responses.
Such an edge offloading of relatively simple AI tasks, or \textit{`Edge Intelligence'} (EI)~\cite{10.1145/3409977,10.1145/3429945}, manages to alleviate the controversy between broadened requirements of modern IAs and the conventional CI paradigm.

The rapid development of EI and corresponding prototypes demonstrates that, due to edge devices' heterogeneous resource constraints, the Cloud is still critical to modern production level IAs with multi-faceted performance requirements~\cite{zhou2019edge}.
Increasing IA developers start to focus on efficiently leveraging edge resources under cloud coordination to collaboratively conduct AI tasks with optimized performance~\cite{Song2018,Abdelzaher2020}, or \textit{`Edge-Cloud Collaborative Intelligence'} (ECCI).
ECCI relies on pivotal interdisciplinary technologies of cloud and edge computing (supporting ECCI infrastructure and runtime), and ML/DL-based AI (introducing rich IA workloads).

Existing ECCI applications (\textit{e.g.}, HOLMES~\cite{10.1145/3394486.3403212} for healthcare,
EdgeRec~\cite{10.1145/3340531.3412700} for E-commerce, SurveilEdge~\cite{Wang2020} for urban surveillance, and general solutions like CLIO~\cite{10.1145/3372224.3419215} and SPINN~\cite{Laskaridis2020}) are individually developed and deployed by either academic researchers or industrial communities, where both the application design and system implementation are highly \textit{developer-dependent} and \textit{scenario-specific}.
For example, SurveilEdge~\cite{Wang2020} is a typical ECCI application for real-time intelligent urban surveillance video query.
In its prototypical implementation, the developers depend on relatively higher edge computation capabilities (\textit{i.e.}, X86 PCs) to support system scaling without subtly designing an ECC infrastructure management scheme.
For the ease of implementation, they hard-code the load balancing policy with the video query workload for latency reduction.
Additionally, to achieve intelligent video query, the entire solution is specifically designed to support CNN training and inference workloads, where dedicated service links (\textit{e.g.}, message service links) among all application components are individually configured to achieve edge-cloud collaborations.
Without impacting the application performance, such developer-dependent design and implementation, however, are impeding others to migrate the application to general ECC infrastructures (\textit{e.g.}, resource-constrained Industrial IoTs) or pursue customizable performance optimizations (\textit{e.g.}, joint optimization of latency and bandwidth consumption).
Moreover, if others want to adopt SurveilEdge (or other existing applications) as the backbone of other applications, driven by different DL models and deployed at different infrastructures, corresponding DL runtimes and different ECC services have to be designed and implemented by the adopters themselves thoroughly.
Such a non-generic manner is severely hindering the proliferation of production level ECCI applications.

Therefore, \textit{for the cost-efficient implementation of high-performance production level ECCI applications, it is necessary to construct a unified platform handling both ever-increasing edge and cloud resources and emerging IA workloads with increasing scale and complexity.}
Particularly, to construct such a platform, the following \textbf{four challenges} need to be explicitly addressed:

\textbf{Support for unified management of hierarchical and heterogeneous infrastructures.}
The efficient implementation of ECCI applications requires unified management of not only infrastructures offered by traditional centralized cloud providers but also heterogeneous computation, storage, and network resources geographically dispersed at the edge.
The development and deployment of ECCI application components on edge devices are extremely inefficient due to the lack of a unified platform.
Furthermore, it is infeasible to directly migrate IaaS and PaaS technologies in cloud computing to the management of inherently distributed edge resources~\cite{Bagchi2019}.

\textbf{Support for user-transparent ECC services.}
ECCI application developers require services providing user-transparent edge-edge and edge-cloud collaborations.
In most cases, components of existing ECCI applications are independently deployed on edge nodes, only interacting through services deployed on the Cloud.
Such a manner increases both bandwidth cost and response latency.
Few existing edge services (\textit{e.g.}, Dapr~\cite{dapr}) can improve edge autonomy and application performance to a certain extent.
However, due to the lack of links between edge and cloud services, they cannot provide user-transparent collaborative services to developers.

\textbf{Support for complex IA workloads.}
Efficient ECCI application implementations require comprehensive system-level supports to complex IA workloads like ML/DL model training and inference, which cannot be provided by existing cloud and edge computing platforms.
For instance, in edge computing systems for IoT data processing, the message-driven communication solution for transmitting KB-level sensor data cannot effectively handle the transmission of DL models as large as hundreds of MBs.
Moreover, most existing distributed ML/DL solutions are designed for datacenter networks with high bandwidth and low transmission latency.
Such methods are inefficient in ECC systems with inherent constraints like prolonged and unstable End-to-End (E2E) communication latency.

\textbf{Support for unified optimization of ECCI applications.}
Unified performance optimization mechanisms are important to efficient ECCI application implementations.
For most existing edge computing applications, the efficiency of resource utilization highly depends on the developer's design, where effective optimizations require a profound understanding of system architectures and optimization theories~\cite{Harchol2019}.
For existing ECCI applications, except for the multi-component development and cross-device deployment of inherently complex IA workloads, the developers also have to deal with the overall performance optimization across ECC infrastructures by themselves, not to mention the difficulties in application debugging, monitoring, and profiling caused by the distributed and heterogeneous environment.
Such a requirement is quite challenging to not only developers of emerging ECCI applications, but also those who want to migrate existing IAs to ECC infrastructures.

\section{ECCI Application Patterns} \label{sec:apppattern}
Currently, there exists no commonly-accepted abstraction of general ECCI application patterns, which are critical to improving the efficiency of ECCI application development and deployment.
As the foundation of the unified platform, considering the subject of different application tasks, we extract \textbf{four} common patterns, \textit{i.e.}, \textit{ECC processing}, \textit{ECC training}, \textit{ECC inference}, and \textit{hybrid collaboration}.

\textbf{\textit{ECC Processing}} of data is the basis of various ECCI applications.
Collaborative data processing applications are often built as \textit{pipelines} or \textit{Directed Acyclic Graphs} (DAGs).
For example, the Steel framework~\cite{Noghabi2018} deploys a streaming analytic pipeline of different data processing tasks (\textit{e.g.}, filtering, anomaly detection, and storage) for ECC IoT anomaly detection applications.

\textbf{\textit{ECC Training}} refers to conducting ML/DL model training based on edge-cloud collaborations.
Unlike ECC processing, ECC training requires complex interactive and iterative data and control flows between edges and the Cloud (\textit{e.g.}, training data, model, and hyper-parameter exchanges).
For instance, Federated Learning (FL) is a typical ECC training application, which conducts ML training across devices to protect data privacy (\textit{e.g.}, Gboard Mobile Keyboard~\cite{Google2017} and Apple QuickType Keyboard~\cite{Apple2019}), and to bridge data silos (\textit{e.g.}, model training for bank fraud detection~\cite{WeBank2019}).

\textbf{\textit{ECC Inference}} focuses on ML/DL model inference, where only forward propagation is conducted.
Generally, ECC inference is achieved through either intra-model or inter-model collaborations.
In intra-model solutions, a single DL model is decomposed by layers into two parts (\textit{i.e.}, neural network partitioning) deployed at edges and the Cloud respectively for collaborative inference (\textit{e.g.}, Neurosurgeon~\cite{neurosurgeon}, SPINN~\cite{Laskaridis2020}, and JointDNN~\cite{jointdnn}).
In inter-model ones, however, multiple DL models with different functionality or performance are deployed at edges and the Cloud respectively for collaborative inference (\textit{e.g.}, VideoEdge~\cite{Hung2018} and SurveilEdge~\cite{Wang2020}).

\textbf{\textit{Hybrid Collaboration}} combines at least two of three ECCI application patterns above or integrates additional CI/EI capabilities into ECCI applications.
For example, ShadowTutor~\cite{Chung2020} enables robust HD video semantic segmentation with significant throughput improvement and network data transmission reduction.
Here, cloud servers conduct both the inference of the heavy and general `teacher' model and the training of the lightweight `student' model.
Mobile edge devices conduct the 'student' model inference.

\section{ECCI Platform Design Principles} \label{sec:designprinciple}
In this article, we aim to construct a unified platform for the efficient development and deployment of ECCI applications.
It is required to provide efficient management of heterogeneous ECC infrastructures, user-transparent ECC services, and customizable performance optimizations, supporting scalable, reliable, and robust ECCI application development and deployment.
The desired platform should be treated as \textit{ECCI-as-a-Service} (ECCIaaS), similar to the concept of Machine Learning-as-a-Service (MLaaS).
Particularly, we extract \textbf{five} essential design principles as follows.

\textit{\textbf{Principle One:} an instance of ECCI application should be an integrated entity that can be managed in a scalable manner.}
This principle requires the unified management of typical edge and cloud infrastructures, including hardware nodes like edge gateways, clusters like Kubernetes~\cite{kubernetes}, virtual machines, and third-party cloud services like Azure IoT Hub~\cite{iothub}.
Any operation of ECCI applications (\textit{e.g.}, deployment and monitoring) should be carried out on large-scale collaborative infrastructures organized as a unity.
ECCI applications should be able to provide continuously available services when the infrastructures are scaled or upgraded.

\textit{\textbf{Principle Two:} ECCI application components at edges and the Cloud should be able to operate in both collaborative and autonomous manners.}
Unlike the datacenter network on the Cloud, the edge-cloud network has limited capabilities (\textit{e.g.}, bandwidth), and may perform unstably.
While supporting collaborations with the Cloud, edges should be able to cache data and provide partial services autonomously to mitigate the impact of network partitioning.

\textit{\textbf{Principle Three:} orchestration is essential to ECCI applications.}
Except for edge-cloud separations, modularized ECCI application components have specific requirements of computation and storage resources, as well as deployment locations.
Moreover, there can be multiple applications co-located at the same infrastructure.
Therefore, component orchestration is necessary to ensure that all applications' resource and user requirements can be satisfied.

\textit{\textbf{Principle Four:} provide in-app control of ECCI applications.}
In most cases, offloading computation to edges may not directly improve application performance.
Here, in-app control optimization has been demonstrated to be effective in various aspects like bandwidth saving~\cite{nigade2020} and E2E latency reduction~\cite{Ren2020}, which should be seriously considered for application performance enhancement.

\textit{\textbf{Principle Five:} support as many types of ECCI application workloads as possible.}
ECCI application scenarios are ever-increasing, such as federated model training and ECC model inference.
It is essential for the platform to support common application patterns and services, facilitating efficient development and deployment of a broadened spectrum of ECCI applications.

\section{Application-Centric ECCI Platform} \label{sec:ace}
Driven by all principles above, the explicit design of our Application-Centric ECCI (ACE) platform is as follows.

\begin{figure*}[htbp]
  \centering
  \includegraphics[width=0.85\textwidth]{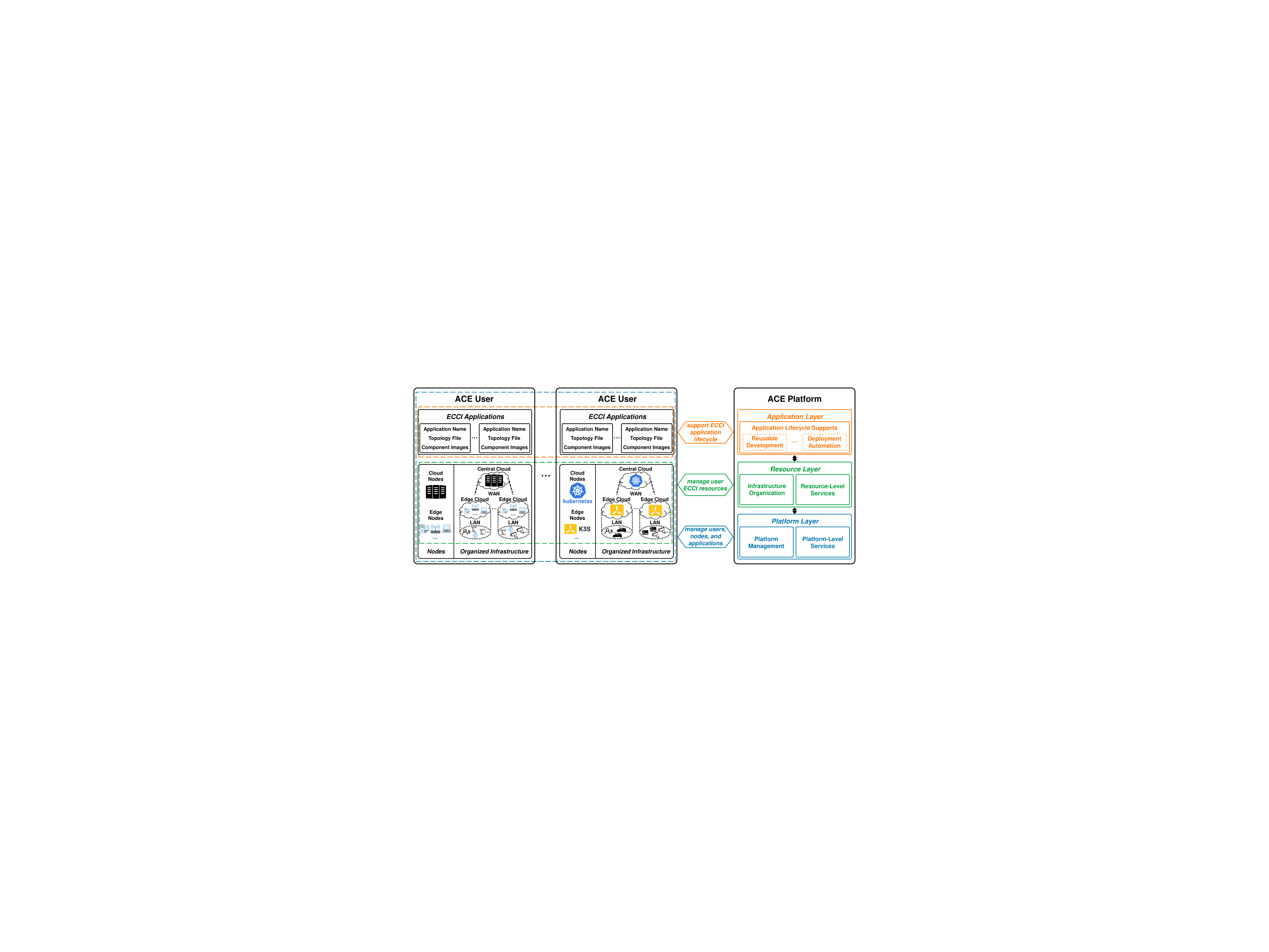}
  \caption{The general architecture of ACE.} 
  \label{fig:design}
\end{figure*}

\subsection{Overview} \label{subsec:aceoverview}
We illustrate the general ECCI application development and deployment procedure based on ACE in Figure~\ref{fig:design}.
For application developers, this procedure comprises \textbf{three} major phases, \textit{i.e.}, \textit{user registration}, \textit{application development}, and \textit{application deployment}.

In the \textit{user registration} phase, any ECCI application developer can register at ACE as a \textit{platform user}.
The user first requests the registration of an ECC infrastructure at ACE, and registers all his/her edge and cloud \textit{nodes} to form an infrastructure according to operational instructions replied by ACE (see Part~\ref{subsubsec:infraorg}).
Here, a \textit{node} can be either a physical device or a virtual service like an edge gateway, a cloud server, a private or public cloud, \textit{etc.}
The user can also select to deploy different resource-level services based on service components provided by ACE on the infrastructure, which can be shared among all his/her ECCI applications (see Part~\ref{subsubsec:resourceservice}).

Then, in the \textit{application development} phase, the user implements applications in a modularized manner.
Specifically, for each application, different components are separated according to user-defined business logic or functionalities.
Meanwhile, requested by ACE, the user deliberately decouples application control flows with workload flows for collaboration optimization and component reuse (see Part~\ref{subsubsec:reusedev}).
All components are then implemented using the ACE Software Development Kits (SDKs), and encapsulated into containers that can be deployed on edge or cloud according to components' resource and user requirements.
For each application, the user constructs a \textit{topology file} describing component relations and resource and user requirements of each component.
All component images and corresponding topology files are then submitted to ACE.

Finally, in the \textit{application deployment} phase, ACE determines a \textit{deployment plan} for all components of a specific application according to the topology file, guaranteeing that all resource and user requirements are satisfied (see Part~\ref{subsubsec:depauto}).
According to the plan, the application can be deployed on the user's ECC infrastructure through ACE.
All deployed applications are continuously monitored by ACE for maintenance, and corresponding users can upgrade, monitor, and remove their applications at any time.

To achieve the procedure above, we construct our ACE platform in a hierarchical manner with \textbf{three} layers, \textit{i.e.}, \textit{platform layer}, \textit{resource layer}, and \textit{application layer}.
The general architecture of ACE is illustrated in Figure~\ref{fig:design}.
Details of each layer are as follows.

\subsection{Platform Layer} \label{subsec:aceplatformlayer}
This layer manages the ACE platform, all registered users, and users' nodes and applications.
It also offers platform-level services for users and their applications.

\subsubsection{Platform Management} \label{subsubsec:platformmanagement}
Our platform-layer manager comprises \textit{controller}, \textit{orchestrator}, \textit{API server}, \textit{Pub/Sub service}, \textit{monitoring service}, and \textit{user interfaces}:

\textbf{Controller} 
manages platform users, their nodes and applications, \textit{e.g.}, registers and deletes users, shields failed nodes, and controls node component deployment.

\textbf{Orchestrator}
determines a deployment plan for all components of each application based on the topology file (see Part~\ref{subsubsec:depauto}), ensuring resource (\textit{e.g.}, computing) and user (\textit{e.g.}, location) requirements of all components are satisfied.

\textbf{API Server}
provides uniform APIs for querying and manipulating the status of ACE entities (\textit{e.g.}, users, nodes, applications) to other platform manager components (\textit{e.g.}, orchestrator, controller).

\textbf{Pub/Sub Service}
provides a bi-directional data transmission channel between ACE and user nodes and applications (\textit{e.g.},  delivering deployment instructions from the controller to user nodes).

\textbf{Monitoring Service}
collects the status, performance metrics, and runtime logs of ACE, user nodes and applications.

\textbf{User Interfaces}
enhances ACE's user-friendliness with Command Line Interface (CLI) and web-based dashboard.
For example, the dashboard with a `drag-and-drop' visual application editor can be used for handy application development.

\subsubsection{Platform-level Services} \label{subsubsec:platformservice}
Platform-level services are not ACE's internal features.
They can be implemented as requested to improve the efficiency of ECCI application development and deployment based on ACE.
Following are two typical examples:

\textbf{Image Registry}
hosts ACE-provided images (\textit{e.g.}, controller, orchestrator), generic runtime images (\textit{e.g.}, Python runtime), and user-provided customized application images.

\textbf{Validation Testbed}
allows users to develop, debug, and monitor ECCI applications on an SDN-based application validation testbed.
For example, the impact of edge-cloud channel dynamics (\textit{e.g.}, bandwidth, delay, jitter) on the testbed can help users understand the actual performance of an ECCI application in real-world networks.

\subsection{Resource Layer} \label{subsec:aceresourcelayer}
This layer manages the ECC infrastructure of each user.
It also provides resource-level services shared among applications deployed on the same infrastructure.

  \begin{figure*}[htbp]
    \centering
    \includegraphics[width=0.85\textwidth]{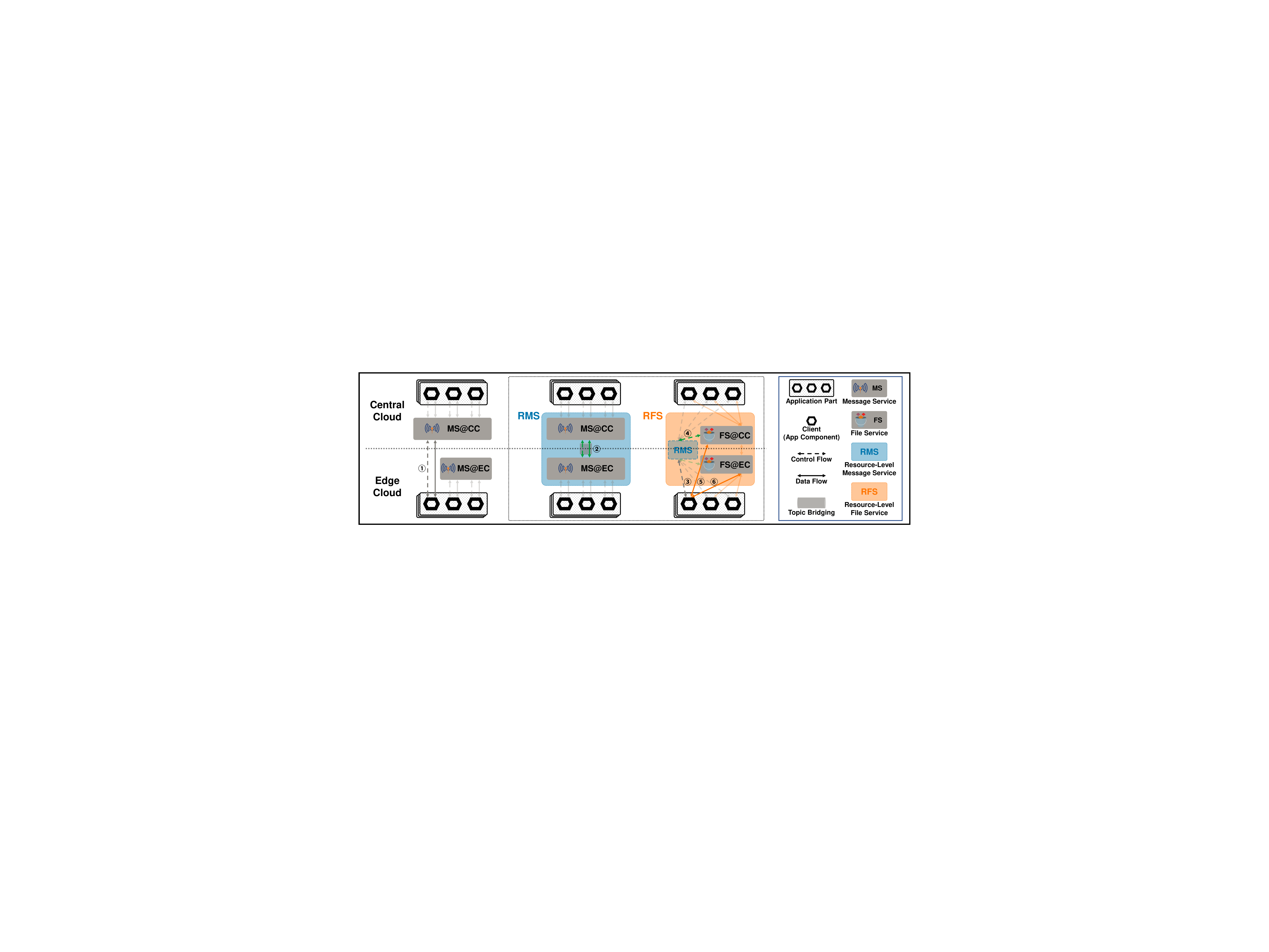}
    \caption{Illustration of ACE provided resource-level services.}
    \label{fig:services}
  \end{figure*}

\subsubsection{Infrastructure Organization} \label{subsubsec:infraorg}
Considering \textit{Principles One} and \textit{Two}, ACE requires all nodes of each user to be organized as several \textit{Edge Clouds} (ECs) and one \textit{Central Cloud} (CC) to host  scalable ECCI applications, and to enable autonomous operations of edge components.
For a specific user, all his/her edge nodes are divided into several groups according to the user's preferences (\textit{e.g.}, in terms of nodes' geographical locations or resources).
Each group is organized as an EC, serving all end nodes (\textit{e.g.}, IoT sensors and cameras) that can access the EC through Local Area Network (LAN).
All cloud nodes of the user are organized as a single CC, and it can interact with each EC through Wide Area Network (WAN).
For each EC and the CC, internal nodes are organized as a cluster (similar to Kubernetes ideally, or a node pool for simple implementation).

Treating each EC and the CC as a resource-level operational unit allows ACE to effectively manage the infrastructure and deploy applications on such an infrastructure, without considering the explicit management of potentially massive underlay nodes.
Moreover, when there is no cloud coordination caused by either CC or edge-cloud communication failure, each EC as a cluster remains (partially) functional, enabling local area collaborations among edge components based on corresponding edge services.

Receiving the user's registration request, ACE assigns a unique infrastructure ID to the user, and establishes a node information record for infrastructure organization.
Meanwhile, ACE assigns a unique second layer ID affiliated to the infrastructure ID to each EC and the CC claimed by the user, where corresponding node registration instructions are generated automatically.
Replied from ACE, such instructions are executed by the user on nodes.
An agent is deployed on each node, informing ACE of the node information and the EC or CC the node belongs.
ACE assigns a unique third layer ID affiliated to the EC or CC's ID to each node.
The agent is also used for application deployment and application status collection.

\subsubsection{Resource-level Services} \label{subsubsec:resourceservice}
For ECCI applications with the typical patterns discussed in Section~\ref{sec:apppattern}, essential services like small/big packet communication and data caching/storage are commonly required~\cite{Harchol2019,MongaRS2019,LiL2020}.
In a specific ECC infrastructure, existing services supporting ECCI applications are conventionally deployed on both ECs and the CC, serving EC and CC clients (\textit{i.e.}, application components) respectively to ensure the autonomy of ECs.
Each service is accessible to all applications deployed on the same infrastructure.
However, due to the lack of links between edge and cloud services, conventional services require application developers to handle complex edge-cloud interactions.
Treating conventional message service for small packet communication as an example, as shown in Figure~\ref{fig:services}, for edge-cloud unicast communications, all EC clients have to directly access the message service at CC (\textit{i.e.}, \circled{1}) to communicate with CC clients.
Here, the developer has to handle the CC message service authorization to each EC client individually, which is quite expensive for large-scale ECCI applications.

Considering \textit{Principle Five}, to facilitate efficient application development, ACE prefers to provide E2E resource-level services with unified interfaces to EC and CC clients, respectively.
Therefore, \textit{long-lasting links} between services on ECs and the CC need to be established.
Some conventional services support the direct establishment of such links (\textit{e.g.}, service bridging for the message service).
Specifically, as shown in Figure~\ref{fig:services}, ACE implements a \textit{resource-level message service}, where the long-lasting link between EC and CC message services (\textit{i.e.}, \circled{2}) is established using MQTT topic-bridging~\cite{roger_mosquitto}.
Here, edge-cloud interactions are conducted by ACE provided SDK, and each client only needs to interact with its local service with a dedicated interface.
For other services, directly establishing long-lasting links is expensive or even infeasible.
For example, the link between edge and cloud file services could be established using file synchronization, which induces additional requirements on network condition, computation, and access authorization.
Instead, ACE uses the resource-level message service to establish long-lasting links for other services.
ACE implements a \textit{resource-level file service}, whose control flow (\textit{e.g.}, \circled{3},\circled{4}) is separated from the data flow and handled by the resource-level message service.
Furthermore, the proverbial object storage service is used to handle the data flow (\textit{e.g.}, \circled{5},\circled{6}) for transmission simplification.

Note that, as shown in Figure~\ref{fig:services}, three types of links are used in resource-level services, \textit{i.e.}, ad-hoc links (grey) for repetitive interactions, ad-hoc links (orange) for one-off interactions, and long-lasting links (green) established once the service is deployed.
Besides, resource-level services should provide basic operations for applications through their lifecycle (\textit{e.g.}, temporary storage for intermittent models and data, and permanent storage for final trained models in the file service).

\subsection{Application Layer} \label{subsec:aceapplicationlayer}
This layer supports user applications through the entire lifecycle.

\subsubsection{ACE Supported ECCI Application Lifecycle.} \label{subsubsec:applife}
As a unified platform, ACE supports each application through its entire lifecycle (\textit{i.e.}, designing, coding, building, testing, deploying, and monitoring).
For designing, ACE provides a standard specification (\textit{i.e.}, the topology file) to achieve modularized development for ACE-organized ECC infrastructures.
For coding, ACE provides the SDKs with access to resource-level services for application components and the user interface to access the essential Integrated Development Environment (IDE).
For building, ACE provides the image registry for efficient image management and distribution.
For testing, ACE provides the validation testbed for application verification and evaluation.
For deployment, ACE provides the orchestrator and the controller for automatic deployment.
For monitoring, ACE provides the monitoring service collecting the status of application components and nodes where they are deployed.
Such supports from ACE enable users to develop and deploy basic ECCI applications efficiently.
For applications with specific performance requirements (\textit{e.g.}, the minimal E2E latency), or with advanced architectures (\textit{e.g.}, large-scale components with complex topology), ACE provides \textbf{two} extra supports, \textit{i.e.}, \textit{reusable development} and \textit{deployment automation}.

\subsubsection{Reusable Development.} \label{subsubsec:reusedev}
Considering \textit{Principles Four} and \textit{Five}, ACE requires developers to decouple and separate control and workload planes of all application components.
The control plane conducts in-app control operations, component monitoring, and policy execution (\textit{e.g.} decide the best partition point for intra-model inference solutions~\cite{neurosurgeon,jointdnn}).
The workload plane conducts computation, storage, and transmission instructed by the control plane (\textit{e.g.}, deep feature compression module ~\cite{ChoiB18} or hybrid collaboration pipeline for data processing and inference tasks~\cite{colliflow}).
Such a separation allows ACE to construct a reusable in-app controller, enabling developers to concentrate on implementing ECCI workloads and efficiently contribute to the ACE based ECCI ecosystem.

For the reusable in-app controller, ACE constructs a series of general in-app control operations (\textit{e.g.}, start, filter, aggregate, and terminate), component monitoring operations, and a basic control policy.
Determined by the ECC infrastructure, the controller is constructed at the resource level in an ECC manner (see Part~\ref{subsubsec:infraorg}).
The CC controller conducts global coordination related operations, and the EC controller coordinates components within the EC.
Resource-level services support interactions between CC and EC controllers.
For applications with specific performance requirements, developers can inherit the general in-app controller and override optimization methods as advanced control policies (\textit{e.g.}, the rate control based optimal edge-cloud bandwidth allocation ~\cite{AlvarB21}).

\begin{figure*}[!htbp]
    \centering
    \includegraphics[width=0.85\textwidth]{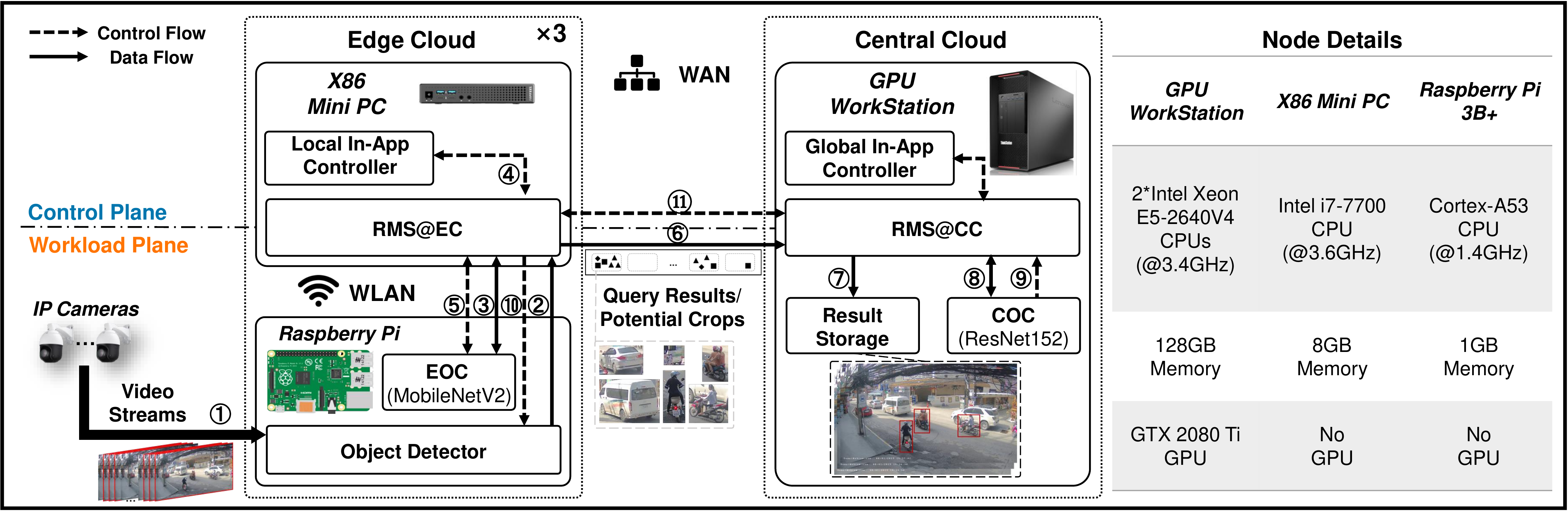}
    \caption{ACE based intelligent video query workflow.}
    \label{fig:video_query}
\end{figure*}

\subsubsection{Deployment Automation.} \label{subsubsec:depauto}
Considering \textit{Principle One}, ACE needs to support efficient application deployment regardless of the topology complexity and the infrastructure scale.
To achieve this, ACE constructs an automatic application deployment method only requiring the application topology file containing information like application specification, component clarifications, parameters, relations, and deployment requirements.
Such a manner prevents users from handling complex component-infrastructure mapping.

Specifically, to deploy an application, the user submits the topology file through the user interface to ACE, and triggers the orchestration process.
According to component deployment requirements, the ACE platform-layer orchestrator binds each component with specific nodes in the infrastructure and resource-level services required, generating the deployment plan.
When the user triggers the deployment process, the ACE platform-layer controller generates the instruction to deploy each component instance on the corresponding node according to the deployment plan, and sends the instruction to the node agent for execution. 
Note that users can manage applications (\textit{e.g.}, update and delete) by modifying the topology file.
For example, for updates, submitting an updated topology file, the user can trigger a thorough update, \textit{i.e.}, ACE deletes the previous application and repeats the entire deployment process.
An incremental update can also be triggered, \textit{i.e.}, ACE automatically deploys updated components according to the new topology file.

\section{How It Works: Intelligent Video Queries using ACE} \label{sec:casestudy}
To validate our platform in supporting efficient and high-performing ECCI application development and deployment, we first present the entire development and deployment process of an intelligent video query application based on ACE, then compare the performance of the application implemented with ACE, CI, and EI, respectively.

\subsection{Application Development and Deployment} \label{subsec:appdevdep}
Video query~\cite{Wang2020,reducto} is one of killer ECCI applications.
To fulfil latency-sensitive user-specific video query requests (\textit{e.g.}, query about the existence of a type of objects in video streams from a geographic area), it generally uses edge and cloud resources to retrieve targeted objects from the video streams with a proper tradeoff between query accuracy and response latency under practical edge-cloud bandwidth limitations.
Based on Subsection~\ref{subsec:aceoverview}, we developed and deployed a video query application (based on ~\cite{Wang2020}) using ACE.

\subsubsection{User Registration.} \label{subsubsec:caseuserreg}
As an ACE user, we first mounted all our nodes and conducted the organization of our ECC infrastructure instructed by ACE.
Our infrastructure comprised a CC (one node, \textit{i.e.}, a GPU workstation), and three ECs (each with four nodes, \textit{i.e.}, an X86 mini PC and three Raspberry Pis).
Node details are in Figure~\ref{fig:video_query}.
For each EC, all edge nodes connected to an individual 100Mbps WLAN.
Each EC connected to CC through WAN (campus network) with software-limited bandwidth (\textit{i.e.}, 20Mbps uplink and 40Mbps downlink) and one-way delay (\textit{i.e.}, 0ms and 50ms as ideal and practical networks, respectively).
Let each Raspberry Pi receive the real-time video stream from a surveillance camera.
We deployed the resource-level message service on the infrastructure.

\subsubsection{Application Development.} \label{subsubsec:caseappdev}
Our application~\cite{ace2022} aimed at fulfilling user-specific video query requests accurately and rapidly through edge-cloud collaborations under practical network limitations (\textit{i.e.}, bandwidth and delay).
We developed the following components:
Data Generator (DG) providing the real-time video stream to the edge node,
Object Detector (OD) rapidly extracting video frame crops potentially containing moving objects from the video stream,
Edge Object Classifier (EOC) conducting lightweight query-specific binary object classification,
Cloud Object Classifier (COC) conducting accurate multi-class object classification,
In-app Controller (IC) executing the control policy,
and Result Storage (RS) saving all query results.
OD on edge nodes was implemented using frame differencing~\cite{Wang2020} (\textit{i.e.}, cropping regions with salient pixel differences across frames) instead of accurate but complex object detector like YOLOv3~\cite{yolov3} for rapid crop extraction on resource-limited edge nodes.
COC on CC was a ResNet152~\cite{resnet} pre-trained on ImageNet ILSVRC15~\cite{ILSVRC15} (4.49\% Top-5 error rate).
EOC was a MobileNetV2~\cite{mobilenetv2} rapidly trained on-the-fly by CC whenever there were user-specific queries.
To form its query-specific training set, video frame crops containing different classes of objects were extracted on CC by a YOLOv3 pre-trained on COCO~\cite{coco} (57.9\% mAP measured at 0.5 IOU) from historical video data uploaded by cameras at (or nearby) the queried area at leisure time, then labelled by COC.
The trained EOC (training details are in ~\cite{Wang2020}) was then deployed on edge nodes in the queried area.
We used real video clips from Youtube Live~\cite{youtubelive} ($30$ fps, $1920 \times 1080$ resolution, various durations) as historical video data and real-time video streams to query.
For a motorcycle query task, EOC's training set had 14,000 crops extracted from clips (170 hours total duration) from 14 surveillance cameras at or nearby the queried area (\textit{i.e.}, historical video data).
Another 6433 `motorcycle' and 68749 `non-motorcycle' crops were extracted as EOC's test set, where EOC achieved 11.06\% error rate under 80\% object identification confidence, tending to be less accurate than COC.
Another three video clips with $5$ minutes duration were used as real-time video streams.
Each node in the three ECs had one of the three clips.

The video query workflow after EOC's deployment is shown in Figure \ref{fig:video_query}.
For each edge node receiving the real-time video stream from DG (\textit{i.e.}, \circled{1}), OD selected three consecutive frames with a specific interval (\textit{e.g.}, 0.5 seconds), and rapidly extracted crops potentially containing moving objects.
Such crops were classified by EOC (\textit{i.e.}, \circled{2},\circled{3}), and the results were used by IC for crop scheduling based on the \textit{Basic Policy} (BP) (\textit{i.e.}, \circled{4},\circled{5}).
If the object identification confidence of a crop was above $80\%$, a targeted object was identified (predicted as positive due to the lack of ground truth of the real-time video), and its metadata were sent to RS (\textit{i.e.}, \circled{3},\circled{6},\circled{7}).
If the confidence was below $10\%$, the crop was dropped.
Otherwise the crop was sent to COC (\textit{i.e.}, \circled{3},\circled{6},\circled{8}).
If the Top-5 classification results of the crop on COC contained the targeted label, a targeted object was identified (\textit{i.e.}, predicted as positive), and its metadata was sent to RS (\textit{i.e.}, \circled{8},\circled{7}).
Since BP may induce queue backlog at EOC and frequent reprocessing at COC, we constructed an \textit{Advanced Policy} (AP) (\textit{i.e.}, \circled{4},\circled{10}) based on BP as a customized IC to further reduce E2E Inference Latency (EIL).
AP collected and estimated EILs of EOC (\textit{i.e.}, \circled{5},\circled{4}) and COC (\textit{i.e.}, \circled{9},\circled{11},\circled{4}) to guide crop uploading from OD (\textit{i.e.}, load balancing~\cite{Wang2020}, always sent to the one with a lower estimated EIL, \circled{2}, \circled{6},\circled{8}).
Then, AP reduced crops uploaded from EOC to COC by shrinking the identification confidence thresholds when either EOC's or COC's EIL got deteriorated.

\begin{figure}[tbp]
  \centering
  \includegraphics[width=0.85\columnwidth]{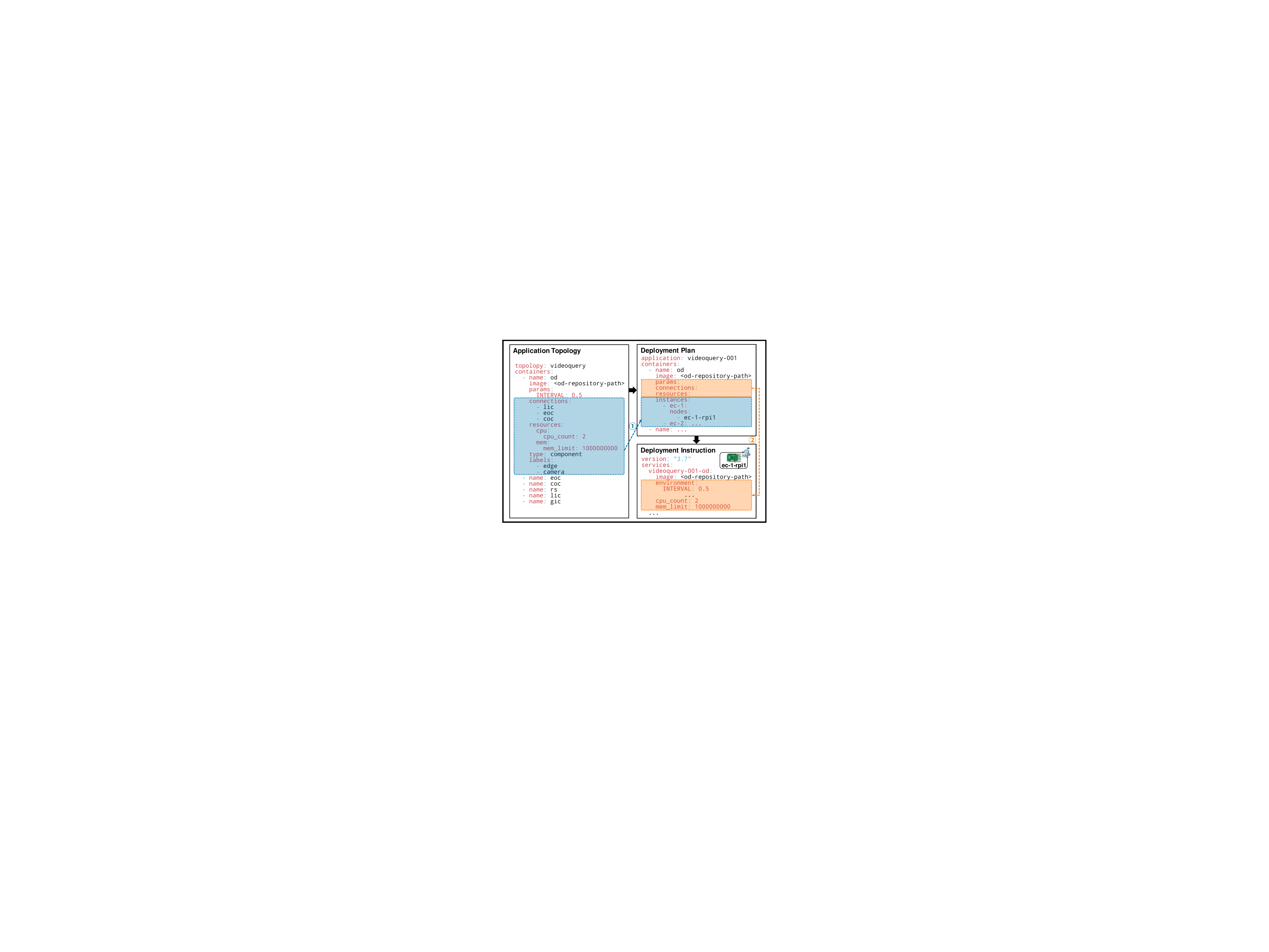}
  \caption{Automatic application deployment.} 
  \label{fig:deployment}
\end{figure}

\subsubsection{Application Deployment.} \label{subsubsec:caseappdep}
As shown in Figure~\ref{fig:deployment}, we submitted a topology file to ACE, which was an extended YAML file containing meta information of both the application and all components.
We illustrate the deployment of component OD as an instance.
Receiving the topology file, as Step \circled{1}, the orchestrator determined the node(s) (\textit{e.g.}, Raspberry Pi `ec-1-rpi1' on edge cloud `EC-1') satisfying all requirements of OD (\textit{i.e.}, `connections' implying OD's dependencies with components Local In-app Controller (LIC), EOC, and COC, `resources' implying CPU and memory required by OD, and `labels' implying that OD should be deployed on edge nodes connected to cameras).
Such decisions were recorded in the deployment plan (\textit{i.e.}, `instances'), a topology replica modified by the orchestrator.
Note that, to manage nodes in an EC as a cluster, ACE can delegate node-level orchestration to the EC.
Receiving the deployment plan, as Step \circled{2}, the controller transformed information of OD instances into specific deployment instructions in a standard Docker-compose YAML file, which was distributed to the node agent (\textit{e.g.}, the container engine at `ec-1-rpi1') for OD deployment.

\subsection{Impact of Implementation Paradigm on Intelligent Application Performance} \label{subsec:caseeva}
\begin{figure}[tbp]
\centering
\includegraphics[width=0.85\columnwidth]{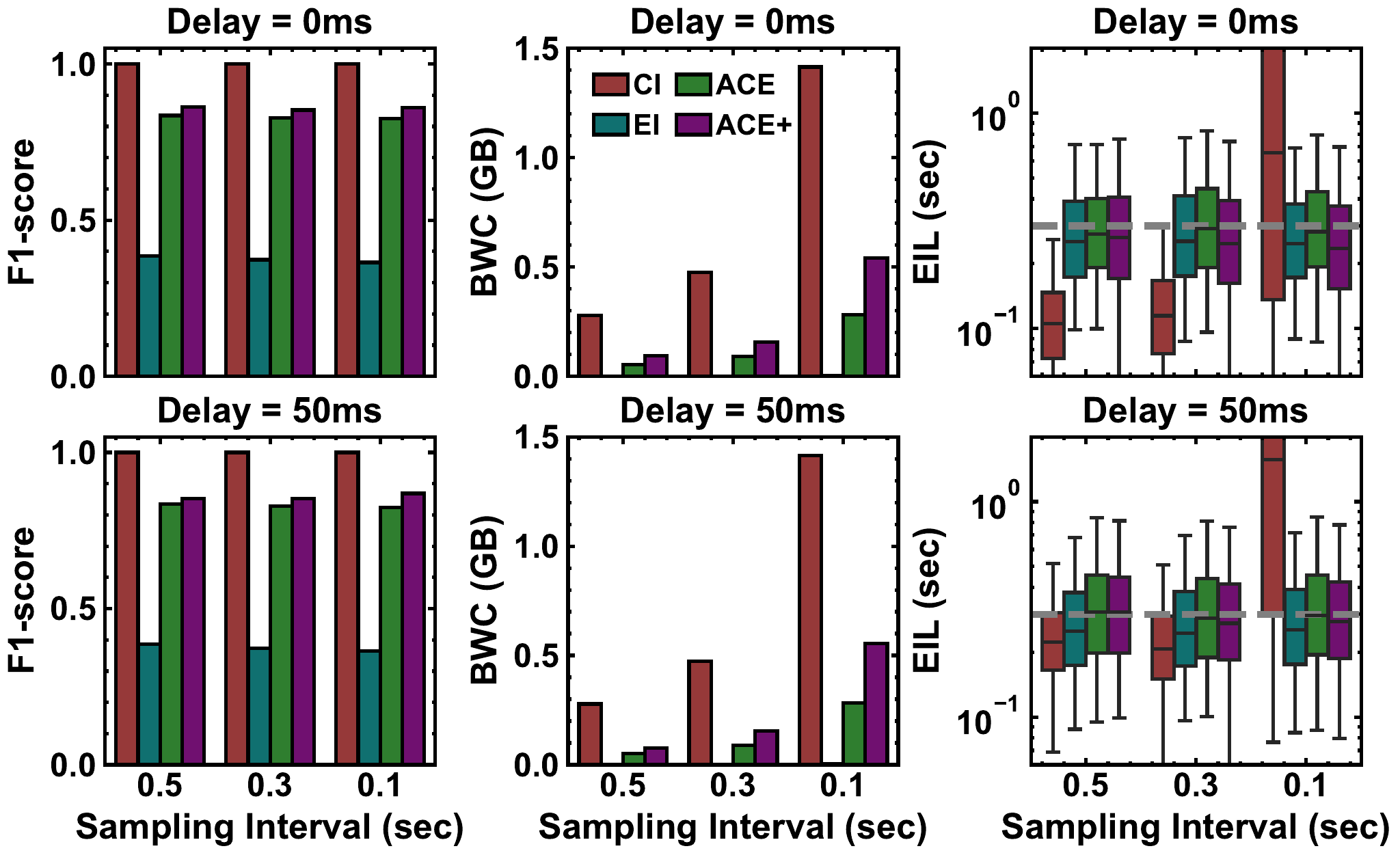}
\caption{Intelligent video query performance.} 
\label{fig:exps}
\end{figure}

We compared the performance of our application implemented with different paradigms.
For CI, COC was deployed on CC. 
For EI, EOCs were deployed on ECs.
For ECCI, based on ACE, two versions of the application with BP (ACE) and AP (ACE+) were deployed.
Different system loads were simulated by varying the sampling interval of frame differencing in OD from 0.5 to 0.1 seconds.
Since all comparatives used the same OD, we compared their video query performance using their object classification performance.
Particularly, we used \textit{F1-score}~\cite{f1score}~\footnote{
Since real-time video streams to query were not labelled, we classified all crops extracted by OD during the entire query task with COC after the task was finished, and treated COC's predicted labels as the query ground truth for F1-score calculation.}, \textit{edge-cloud BandWidth Consumption} (BWC), and \textit{E2E Inference Latency} (EIL)~\footnote{ Time from a crop is transmitted by OD to its predicted label is given by EOC or COC.} as evaluation metrics.
We conducted the motorcycle query task under different system load and edge-cloud network delay (\textit{i.e.}, 0ms for ideal and 50ms for practical) settings.
Results are illustrated in Figure~\ref{fig:exps}.

When the system load increases, F1-scores of CI and EI basically remain the same, where CI using COC only and EI using EOC only achieve the highest and lowest F1-scores under all system loads, respectively.
ACE and ACE+ using COC and EOC collaboratively manage to achieve F1-scores slightly lower than CI but significantly higher than EI.
Unlike EI, in ACE and ACE+, many crops that cannot be confidently classified by EOCs (with a confidence from 10\% to 80\% and dropped by EI) are uploaded to COC.
Compared with CI, few crops are dropped by EOCs (with a confidence below 10\%) in ACE and ACE+.
Besides, the higher the system load, the better ACE+ performs compared with ACE.
Under higher system loads, more crops are directly uploaded from ODs to COC by IC with AP for load balancing in ACE+, reducing crops dropped by IC with BP in ACE.
Furthermore, when the system load increases, ACE+ achieves higher F1-scores under practical than ideal network delay.
In ACE+, under practical network delay, fewer crops are uploaded from EOCs to COC to avoid higher EILs by shrinking the confidence thresholds, and more are from ODs to COC for load balancing.

When the system load increases, BWCs of all except for EI increase.
ACE and ACE+ induce significantly lower BWCs than CI since considerable objects are identified by EOCs.
Furthermore, the higher the system load, the higher BWCs of ACE+ compared with ACE.
In ACE+, some crops (increase with system load) are directly uploaded by IC with AP for load balancing, where, however, only some of them are uploaded by IC with BP in ACE (with identification confidence from 10\% to 80\%), inducing higher BWCs.

When the system load is low, CI induces the lowest EIL under different network delay settings benefiting from COC's fast processing (\textit{i.e.}, the inference time of COC is about 32.3ms on CC, and that of EOC on edge node is above 44ms).
When the system load increases, different from EI, ACE, and ACE+, CI's EIL increases significantly due to the large queue backlog aggregated from all ODs (normal in large-scale edge-cloud systems).
Besides, the practical network delay also enlarges CI's EIL more obviously (significantly higher than the 50ms network delay).
Compared to CI, EILs of EI, ACE, and ACE+ are not obviously impacted by both system load (\textit{i.e.}, low queue backlog at EOCs) and network delay (no/low uploading).
ACE's EIL is slightly higher than EI since EOCs manage to identify most objects, and only a few crops are uploaded to COC.
Furthermore, the higher the system load, the lower EIL of ACE+ compared with ACE.
Some crops (increase with system load) are directly uploaded to COC for load balancing by IC with AP in ACE+.

Compared with CI and EI, ACE-based video query manages to better fulfill query requests accurately and rapidly with efficient bandwidth consumption.
ACE also facilitates developers for customized optimization (\textit{i.e.}, EIL reduction with customized AP).

\section{Future of ACE} \label{sec:discussion}
As a prototype for cost-efficient ECCI application development and deployment, ACE is still in its infancy.
The construction of ACE reveals fundamental challenges to address and sheds light on the vision of an ACE-based ECCI ecosystem deserving explorations.

\subsection{Challenges} \label{subsec:challenge}
\textit{Agile ECCI application orchestration} is critical, but challenging, to improve the performance of ACE-based applications.
ACE's platform-layer orchestrator manages to allocate application components to proper nodes satisfying basic (node-level) resource and user (\textit{i.e.}, edge/cloud deployment) requirements.
However, fine-grained orchestration under more explicit constraints is still hard to achieve, which is significant to fully infrastructure utilization.
Furthermore, ACE's static application orchestration cannot adjust to application or infrastructure changes. A dynamic orchestrator is also necessary.

\textit{Resource contention prevention} has to be further investigated to ensure the performance of ECCI applications co-located at the same infrastructure.
Currently, ACE manages to achieve component-level resource isolation through containerization, and support inter-component resource allocation optimization through the customized in-app controller, where, however, application-level resource isolation and allocation is still an open issue.
Critical resources like edge-cloud bandwidth should be allocated appropriately to co-located applications under ACE's coordination.
It is also promising to integrate the serverless technology~\cite{Castro2019} for elastic resource allocation that cannot be directly achieved by container-based solutions.

\textit{Security} is another critical issue.
ACE now contains no security module, where state-of-the-art encryption and authentication techniques can be directly integrated for fundamental secrecy.
The actual challenge, however, is access control.
In our design, an ACE user has full access to his/her infrastructure and ECCI applications, where no user collaboration is currently supported.
For specific applications (\textit{e.g.}, federated learning) that have to be developed and deployed by multiple users collaboratively, ACE is required to provide a fine-grained access control mechanism.
It needs to ensure that each collaborator has limited access to the shared application and infrastructure without jeopardizing others' privacy.

\subsection{Vision} \label{subsec:vision}
\textit{ACE demonstrates the potential in supporting closed-loop DevOps of ECCI applications.}
ACE manages to facilitate the cost-efficient development and deployment of ECCI applications effectively.
Taking a step further, we believe it is viable to integrate proper operation and maintenance modules into ACE, aiming at the close loop of continuous ECCI application development, deployment, monitoring, delivering, and testing.
Such full DevOps supports will enable ACE to act as the foundation of the approaching ECCI ecosystem.

\textit{ACE is promising in promoting a broad spectrum of production level ECCI applications.}
ECCI applications, especially high-performing ones, are difficult to design, develop, and deploy, which hinders such a paradigm from contributing to the rapidly expanding IA market.
ACE manages to provide supports along the entire ECCI application lifecycle, facilitating general users to conduct unified and user-friendly application development and deployment.
Besides, ACE can also ease the migration of existing IAs based on CI and EI to ECCI applications satisfying specific practical requirements.

\section{Conclusion} \label{sec:conclusion}
ML/DL-based IAs with harsher practical requirements cast challenges on conventional CI implementations.
The emerging ECCI paradigm can support proliferating IAs that, however, are currently developed and deployed individually without generality.
We envision systematic designs of a unified platform for cost-efficient development and deployment of high-performing ECCI applications, guiding us to construct the ACE platform handling heterogeneous resources and IA workloads.
Our initial experience shows that ACE manages to help developers and operators along the entire lifecycle of ECCI applications, where customizable optimizations can be conducted efficiently.
Further research is still required, and we discuss both the challenges and visions of the newborn ACE.

\begin{acks}
This work was supported in part 
by the National Key Research and Development Program of China under Grant 2020YFA0713900;
the National Natural Science Foundation of China under Grants 61772410, 61802298, 62172329, U1811461, U21A6005, 11690011;
the China Postdoctoral Science Foundation under Grants 2020T130513, 2019M663726;
and the Alan Turing Institute.
\end{acks}

\bibliographystyle{ACM-Reference-Format}
\bibliography{references.bib}

\end{document}